\documentclass[twocolumn,aps,prd,showpacs,superscriptaddress,preprintnumbers,floatfix]{revtex4}
\usepackage{graphicx}

\def\be{\begin{equation}}
\def\ee{\end{equation}}
\def\bea{\begin{eqnarray}}
\def\eea{\end{eqnarray}}

\begin{document}


\title{Light Bending as a Probe of the Nature of Dark Energy}
   
\author{F. Finelli}
\affiliation{INAF-IASF Bologna, 
via Gobetti 101, I-40129 Bologna - Italy}
\affiliation{INAF, Osservatorio Astronomico di Bologna, 
via Ranzani 1, I-40127 Bologna -
Italy}
\affiliation{INFN, Sezione di Bologna,
Via Irnerio 46, I-40126 Bologna - Italy}

\author{M. Galaverni}
\affiliation{INAF-IASF Bologna, 
via Gobetti 101, I-40129 Bologna - 
Italy}
\affiliation{Dipartimento di Fisica, Universit\`a di Ferrara,
via Saragat 1, I-44100 Ferrara - Italy}
\affiliation{INFN, Sezione di Bologna,
Via Irnerio 46, I-40126 Bologna - Italy}

\author{A. Gruppuso}
\affiliation{INAF-IASF Bologna, 
via Gobetti 101, I-40129 Bologna - 
Italy}
\affiliation{INFN, Sezione di Bologna,
Via Irnerio 46, I-40126 Bologna - Italy}

\date{\today} 

\begin{abstract} 
We study the bending of light for static spherically 
symmetric (SSS) space-times which include a dark energy contribution. 
Geometric dark energy models generically predict a
correction to the Einstein angle written in terms of the distance to the 
closest approach, whereas a cosmological constant $\Lambda$ does not.
While dark energy is associated with 
a repulsive force in cosmological context, its effect on null 
geodesics in SSS space-times can be attractive as for the Newtonian term. 
This dark energy contribution may be not negligible with respect to the 
Einstein prediction 
in lensing involving clusters of galaxies. Strong lensing may therefore  
be useful to distinguish $\Lambda$ from other 
dark energy models.
\end{abstract}


\maketitle

\section{Introduction}
It is still unclear what drives the universe into acceleration 
recently. While a cosmological constant $\Lambda$ is the simplest 
explanation, its value seems completely at odd with the naive estimate of 
the vacuum energy due to quantum effects. 

An alternative to $\Lambda$ is obtained considering 
a dynamical degree of freedom added to the primordial soup. 
Since dynamical dark energy (dDE) varies in time and space, its fluctuations are potentially 
important in order to distinguish it from $\Lambda$ \cite{CDS}.
Another possibility for the explanation of the present acceleration 
of the universe is given by the geometry itself, through a modification of 
Einstein gravity at large distances \cite{DDG}: 
these are known as geometric dark energy (gDE) models. A non vanishing mass for 
the graviton is among these possibilities \cite{GG}.

{\em Cosmological} observations, such as those coming from 
Supernovae, cosmic microwave background (CMB) anisotropies and large scale structure (LSS), have not 
been able to discriminate 
among dDE models yet (see \cite{UIS} for updated constraints and forecasts). 
It is therefore important to explore observational tests at {\em astrophysical}
level for objects which are detached from the cosmological expansion. 


\section{Deflection of Light}

We shall consider a SSS metric in terms of the physical radius 
$r$
\be
ds^2 = - B(r) dt^2 + A(r) dr^2 + r^2 \, d\Omega ^2
\, .
\label{SSS}
\ee
Such a metric with  
\be
B(r) = A^{-1} (r) = 1- {2 G M \over r} - {\Lambda \over 3} r^{2} \,,
\label{SdS}
\ee
describes the Schwarzschild-de Sitter (SdS) space-time, the vacuum solution
of Einstein equations in presence of a
cosmological constant $\Lambda$ (in $c=1$ units, where $c$ is the 
velocity of light). With the above metric, the classic 
general relativistic tests can be computed in analogy with the 
Schwarzschild (henceforth S) textbook case \cite{weinberg}. We shall restrict here to 
light bending, leaving other results for elsewhere \cite{matteo,prep}.

\begin{figure}
\begin{tabular}{c}
\includegraphics[scale=2]{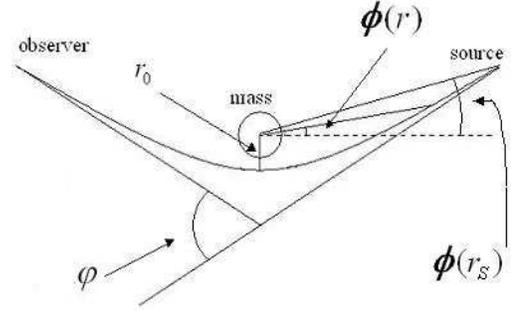}
\end{tabular}
\caption{Deflection of Light}
\label{bend_fig}
\end{figure}

The textbook deflection angle for a photon in a SSS metric can be 
easily extended by keeping into account the finite distance $r$ at which 
the object is located as (see Fig.\ref{bend_fig})
\begin{eqnarray}
\varphi &=& 2 |\phi(r_0)- \phi(r)| 
-\pi \, + 2 \arcsin \left(r_0\over r\right) \, , \nonumber \\
&\equiv&  2 \, I(r,r_0)
-\pi  + 2 \arcsin \left(r_0\over r\right) \, ,
\label{infinito}
\end{eqnarray}
with $r_0$ representing the minimal distance between the geodesic of the 
light and the lens \cite{same,impact} and $I(r,r_0)$ 
given by: 
\bea
I(r,r_0) &=& \int_{r_0}^{r} {A^{1/2}(r^{\prime}) \over r^{\prime}}
\left[ \left(r^{\prime}\over r_0\right)^2
{B(r_0)\over B(r^{\prime})}-1\right]^{-1/2}
\!\!\!\!\!\! dr^{\prime} \nonumber \\
&\equiv& \int_{r_0}^{r} {\cal I} (r^{\prime},r_0) dr^{\prime}
\, .
\label{def_int}
\eea
In the textbook treatment 
\cite{weinberg}, both the distances of the observer and the 
source from the lens are much greater than $r_0$ and the upper limit of 
integration can be taken as $\infty$. For DE metrics with 
non-asymptotically flat terms, it is safe to keep 
finite the upper limit of integration. 

By inserting the SdS metric in ${\cal I} (r,r_0)$, it is easy to check
that the terms including $\Lambda$ cancel out! This cancellation is
due to the particular form of the non-asymptotically flat terms of the SdS 
metric and physically means that $\Lambda$ is truly a potential offset 
for a massless particle (see below). The light bending angle is therefore 
$4 G M/r_0$ even in presence 
of $\Lambda$ \cite{islam,nota}, although all the other 
tests and kinematical quantities
differ from the S case \cite{matteo,prep}.

\section{Geometric Dark Energy Static Spherically Symmetric Metrics}

In order to study the contribution of DE to light bending, we need 
physically motivated SSS metrics which differ from the SdS case. 
As an example, we consider SSS metrics parametrized by:
\bea
\label{metrica:de}
B(r) &=& C - {D \over r} - \gamma_2 \, r^{\alpha} \nonumber \\
A^{-1} (r) &=& C - {D \over r} - \gamma_1 \, r^{\alpha}
\,.
\eea
This parametrization covers SSS metrics of well studied gDE 
models:
\begin{itemize}
\item $C=1\,$, $D=2 G M\,$, $\alpha=3/2\,$, $\gamma_1= 3 \gamma_2/2 = - 
m_g^2 \sqrt{2 G M/13}$ \cite{damour} 
corresponds to the non-perturbative solution found by Vainshtein (V) for a 
massive graviton \cite{vainshtein} with mass $m_g$. 
\textit{Massive gravity} (MG) is an alternative to a cosmological constant 
\cite{GG};
\item $C=1+3\,G M \gamma_1\,$, $D=2 G M + 3 G^2 M^2 \gamma_1\,$, $\alpha=1 
\,, \gamma_1=\gamma_2=\gamma$ corresponds to the 
general SSS in conformal gravity \cite{MK} (see also \cite{DTT} for 
linear correction to the Newtonian potential) . Conformal gravity contains 
the 
SdS solution without adding any cosmological constant to the Weyl action, 
but includes terms linear in $r$ as well. Considering $C \simeq 1$ and 
$D \simeq 2 G M$ is a good approximation for the values of $\gamma$ we are 
interested in;
\item $C=1\,$, $D=2 G M\,$, $\alpha=1/2 \,, \gamma_1=\gamma_2/2 = 
\sqrt{G M/r_c^2}$ 
corresponds to the self-accelerating branch \cite{LS} of the 
brane induced gravity Dvali-Gabadadze-Porrati (DGP) model \cite{DGP}. $r_c$ is the crossover scale beyond which 
gravity becomes five-dimensional.
\end{itemize}
In the parametrization (\ref{metrica:de}) we have omitted a possible $r^2$ 
term, since we have already shown that it does not contribute to
light bending.
For $\alpha = 1$ the SSS metric is valid till to the particle horizon radius
$r_H \sim \gamma^{-1}$ (for the SdS metric this radius is $(3/\Lambda)^{1/2}$). In the other
two cases the metric is roughly limited by the
so-called V radius $r_V$, which denotes the scale at which the
Newtonian term becomes comparable to the non-asymptotically flat term,
i.e. $r_V \sim \left( GM/\gamma \right)^{1/(1+\alpha)}$ with $\gamma ^{-1} = {\rm min}
\left\{ \gamma_1^{-1} \,, \gamma_2^{-1} \right\}$. 
At this scale we expect deviations from Einstein gravity.
\begin{widetext}
We now expand the integrand considering the DE terms in 
Eq. (\ref{def_int}) as second order with respect to the 
Newtonian correction, and, defining $I(r,r_0) = I_{\rm 
E}(r,r_0) + I_{\rm DE}(r,r_0)$ we obtain:
\begin{eqnarray}
I_{\rm E}(r,r_0) = \int_{r_0}^{r} {dr^{\prime}\over r^{\prime}} \left[ 
\left( r^{\prime} \over 
r_0\right)^2-1\right]^{-1/2} 
\left[ 1+ {G M \over r^{\prime}} 
+{G M r^{\prime}\over r_0(r^{\prime}+r_0)} + {3 (G M)^2 \over 2 r^{\prime 2}}+
{3 (G M)^2 \over r_0(r^{\prime}+r_0)} + {3 (G M)^2 r^{\prime 2}\over 2 r_0^2 (r^{\prime}+r_0)^2}\right]
\, , 
\label{IE}
\\
I_{\rm DE}(r,r_0) = \int_{r_0}^{r} {dr^{\prime}\over r^{\prime}} \left[ \left( r^{\prime} \over 
r_0\right)^2-1\right]^{-1/2} 
{\left[ (\gamma_1 -\gamma_2)r^{\prime (\alpha +2)}+\gamma_2 r^{\prime 2}r_0^{\alpha} 
- \gamma_1 \, r^{\prime \alpha}r_0^2 \right]\over {2 (r^{\prime 2}-r_0^2)}}
\, .
\label{IDE}
\end{eqnarray}
It is possible to analitically compute both the terms in Eqs. 
(\ref{IE},\ref{IDE}):
\be
I_{\rm E} (r,r_0) = \frac{\pi}{2} - \arcsin \left(r_0\over r \right) 
+ \frac{GM}{r_0} \left( 1 - 
\frac{r_0}{r} \right)^{1/2} \left[ \left( 1 +
\frac{r_0}{r} \right)^{1/2} + \left( 1 + \frac{r_0}{r} \right)^{-1/2} \right] 
+ {\cal O} \left( \frac{G^2 M^2}{r_0^2} \right) \,,
\ee
\be
I_{\rm DE}(r,r_0) = {r_0^{\alpha} \over 2} I_{\alpha} \left(\frac{r}{r_0}\right)  \left( \gamma_1 - \alpha \gamma_2 
\right) + \gamma_2 {r_0^{\alpha} \over 2} 
\frac{(r/r_0)^\alpha-1}{\sqrt{(r/r_0)^2-1}} \,,
\label{dephi}
\ee
with 
\bea
I_{\alpha} (y) &=& 
\int_1^y d x \frac{x^{\alpha-1}}{\sqrt{x^2-1}} = \frac{\sqrt{\pi}}{2} 
\frac{\Gamma (\frac{1 - 
\alpha}{2})}
{\Gamma (1- \frac{\alpha}{2})}  
+  \frac{y^{\alpha-1}}{\alpha - 1} {_2 F}_1 \left( \frac{1}{2} \,,  
\frac{1-\alpha}{2} \,, \frac{3-\alpha}{2} \,, \frac{1}{y^2} \right) .
\eea
\end{widetext}

It is important to note that the physical structure of gDE metric kill the 
first term in Eq. (\ref{dephi}) for all the three values $\alpha$. 
Of course, the type of DE contribution to the 
deflection angle is not of the parametrized post-Newtonian (PPN) form 
\cite{weinberg}, 
since the correction in the metric coefficient is {\em not} of the PPN 
form. Note that our non-vanishing result for the DGP model corrects 
previous claims in the literature \cite{LUE} and agrees with previous 
result for Weyl gravity when $r \rightarrow \infty$ \cite{EP}.
A non-negligible DE contribution to the total 
deflection angle is expected for large virialized objects, i.e in 
the case of {\em clusters of galaxies}. 

In the following, we split the Einstein term $\varphi_E$ from the dark 
energy correction $\Delta \varphi$ to the total bending angle $\varphi$.
Therefore, the DE contribution has to be compared with the S term assuming 
$r \rightarrow \infty$ 
(up to the second order \cite{second}): 
\begin{eqnarray}
\label{defl:tot}
\varphi_{{\rm E}}&=& \varphi_{}^{(I)}
+\varphi_{}^{(II)} =\nonumber\\
&=&\frac{4 G M}{r_{0}}+2\left(\frac{15 \pi}{8}-2 \right)
\left( \frac{G M}{r_{0}} \right)^{2}
\, .
\end{eqnarray}
This comparison is shown in Table \ref{tab}.

The sign of the DE contribution can be understood by the one dimensional 
motion for the photon:
\be
\frac{1}{2} \left( \frac{d r}{d \lambda} \right)^2 + V_{\rm eff} (r) = 
\frac{1}{2}
\ee
with $d \lambda \equiv r^2 d \varphi/J$ and 
\begin{widetext}
\bea
V_{\rm eff} (r) = \frac{J^2}{2 r^2} \left[ \frac{1}{A(r)} + 
\frac{(\gamma_1 - \gamma_2) r^\alpha}{J^2 B(r)} \right] 
= \frac{J^2}{2 r^2} \left[ 
C - \frac{D}{r} - \gamma_1 r^\alpha +
\frac{(\gamma_1 - \gamma_2) r^\alpha}{J^2 \left( C - \frac{D}{r} - 
\gamma_2 r^\alpha \right) } 
\right] \nonumber \, . 
\eea
This potential leads to the following force $F$ on a photon:
\bea
\frac{F(r)}{J^2} &=& - \frac{d V_{\rm eff}}{J^2 \, d r} = \frac{C}{r^3} 
- \frac{3 D}{2 r^4} + \frac{\gamma_1}{2} (\alpha - 2) r^{\alpha-3} 
 + ( \gamma_2 - \gamma_1) \frac{r^\alpha}{2 J^2 \left( C - \frac{D}{r} -
\gamma_2 r^\alpha \right)^2} \left[ \alpha \frac{C}{r} - 
\frac{D}{2} (1 + \alpha) \right]
\, .
\eea
From the above it is again clear how $\Lambda$ acts as a null force on 
the photon (in agreement with the null contribution in light bending 
\cite{note_mass}). 
When $\alpha < 2$, in addition to the standard terms, the 
third term acts as an {\em attractive force} for $\gamma_1 > 0$. We also 
note that the sign of the fourth term is model dependent \cite{note}.
Although DE in cosmology is associated to a repulsive force (which 
accelerates the expansion), in a static configuration it may add 
to the Newtonian mass term in deflecting light, explaining the 
positive contribution which we find in Eq. (\ref{dephi}).
\begin{center}
\begin{table}[htbp]
\begin{center}
\begin{tabular}{|c|c|c|c|c|c|}
\hline
              & $\varphi_{}^{(I)}$ & $ 
\varphi_{}^{(II)}$ 
& $\Delta \varphi_{\rm MG}$ & $\Delta \varphi_{\rm Weyl}$ & $\Delta 
\varphi_{\rm DGP}$ \\
\hline
Galaxy ($10^{11} \, M_{\odot}$) & $2.0 \times 10^{-6} $ & $1.9 \times 10^{-12} $ & $-4.6 \times 10^{-11} $ & $1.0 \times 10^{-6} $ & $ 3.9 \times 10^{-9}$\\
\hline
Galaxy Group ($10^{13} \, M_{\odot}$) & $1.9 \times 10^{-5} $ & $1.9 \times 10^{-10}$ & $-7.2 \times 10^{-9} $ & $1.0 \times 10^{-5} $ & $2.3 \times 10^{-7} $ \\
\hline
Cluster ($10^{15} \, M_{\odot}$)   & $1.8 \times 10^{-4}$ & $1.8 \times 
10^{-8} $ & 
$-1.1 \times 10^{-6}$ & $ 1.0 \times 10^{-4}$ & $ 6.6 \times 10^{-6}$\\
\hline
\end{tabular}
\end{center}
\caption{Expected deviations of the three theories considered from 
Einstein theory. The parameters used are the following: $m_g = 10^{-31}$ 
eV, $\gamma^{-1} = 10$ Gpc, $r_c= 5$ Gpc. The distance of closest approach are $r_0 = 
10, 10^2, 10^3$ Kpc for galaxy, galaxy groups, clusters, respectively. The 
upper limits of integration have been taken as $r_V$ for $\alpha=1/2 
\,, 3/2$ and $\infty$ for Weyl gravity. We remind that the Hubble 
distance $1/H_0$ is $\sim 4$ Gpc for $H_0 = 72$ km s$^{-1}$ Mpc$^{-1}$.
As written also in the text, the contribution $\Delta\varphi_{\rm BV}$ coming from the region beyond $r_V$ is taken into account only in DGP model \cite{LUE,LS} and it is given by
$\Delta\varphi_{\rm BV}=\left(G(r_{max})-G(r_V)\right)\, 2 G M /r_0$ where 
$G(r)=\left(1-(r_0/r)^2\right)^{1/2}
\left(1+f(\omega)(r_0/r)\right) \left(1+(r_0/r)\right)^{-1}$ 
with $f(\omega)= (\omega +1)/(2 \omega +3) $. The parameter $\omega = -3 r_c H_0$ is set to $-15/4$ and $r_{max}=1 $ Gpc. }

\label{tab}
\end{table}
\end{center}
\end{widetext}

\begin{figure}
\begin{tabular}{c}
\includegraphics[scale=1]{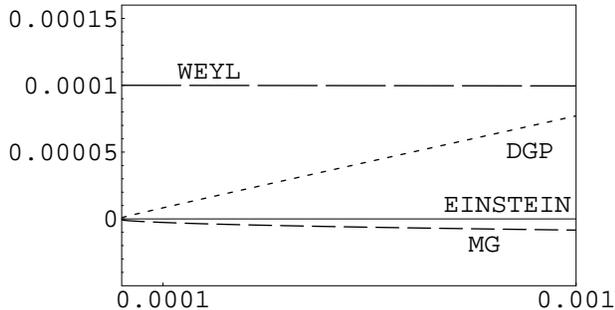}
\end{tabular}

\caption{Dark Energy contribution $\Delta \varphi_{\rm DE}$ to the bending 
angle (in radiants) as a 
function of $GM/r_0$. The solid line ($\Delta \varphi_{\rm DE}=0$) is the 
Einstein prediction for $\Lambda$
; the dotted, long-dashed and dashed are 
the DGP, Weyl and MG models, respectively, with 
$r_0=1$ Mpc (the other parameters are those of Table \ref{tab}). 
Note that $\gamma$ is $M$ independent \cite{MK}.} 
\label{fig}
\end{figure}

We note that the validity of the SSS metrics for $\alpha=1/2 \,, 3/2 \,$ 
up to the V radius - and not up to the particle horizon - can be 
an important limitation to the applicability of our findings. 
In the DGP (massive gravity) model with $r_c = 5$ Gpc ($m_g \sim 10^{-32}$ 
eV), the V radius 
for the Sun is $3.2 \times 10^{18}$ m ($7.5 \times 10^{20}$ m), and 
therefore much larger than the 
size of the solar system $\sim 6 \times 10^{12}$ m 
(taken as the size of the Pluto orbit). For 
clusters with mass $\sim 10^{15} M_{\odot}$, instead, the V 
radius is 10 Mpc for DGP and 24 Mpc for MG 
(with the same parameters used above): such radii are remarkably 
close to the intercluster distance, i.e. ${\cal O} (10) {\rm Mpc}$. 
It is then clear that an understanding of the SSS metrics beyond the 
V scale (the so-called matching problem \cite{damour}) is needed 
for a quantitatively exact calculation of light bending in these models. 
While the MG metric is not known beyond $r_V$, the DGP metric beyond $r_V$ admits a scalar-tensor description \cite{LUE,LS}, 
which we use for the values reported in Table \ref{tab} (see the caption) 
and in Fig.\ref{fig} where we show the prediction of the considered models 
compared with Einstein theory as function of $GM/r_0$.

\section{Discussions and Conclusions} 

We have discussed light bending in SSS with DE. The importance of general 
relativistic tests, such as the 
perihelion precession, has been already emphasized
for the DGP model \cite{DGZ,LS}, while little attention was previously 
paid to light bending. These tests are complementary to the 
observational signatures of dark energy in cosmological context, mainly 
based on the behaviour of perturbations. In cosmology, an important difference 
between $\Lambda$ and dDE or gDE is the presence of DE perturbations 
in the latter case, which are at least gravitationally coupled to the other 
types of matter. Such DE perturbations are therefore a key point 
to distinguish $\Lambda$ from dDE or gDE in CMB and LSS, and sometimes may 
become so important to strongly constrain models \cite{CDS,CF,koyama} with 
respect to what Supernovae data can do.

In this article
we have 
shown that in objects which have 
detached from the expansion of the universe, $\Lambda$ may be 
distinguishable from other DE models through the bending of light.
In order to link our findings with observations, we should insert 
$\varphi$ in the lens equation, e.g. \cite{schneider}:
\begin{equation}
\theta-\beta=\frac{d_{SL}}{d_{OS}} \varphi
\end{equation}
where $\theta$ and $\beta$ are the angular positions of the image 
and of the source measured respect to the line from the observer 
to the lens; $d_{LS}$ and $d_{OS}$ are the angular diameter distances 
between the lens and the source and between the observer and the source, 
respectively.
On considering for simplicity alignment between the lens and the 
source, an Einstein ring forms with angle $\theta_{E} = \theta (\beta = 0)$. 
From our results, $\theta_E$ is affected by 
both the non-perturbative SSS potential around the lens 
($ \varphi \ne 4 G M/r_0$ if DE $\ne \Lambda$) and the 
cosmology of a given model. 
The gDE corrections to the Einstein deflection
angle for clusters in Eq. (\ref{dephi}) are as important as the cosmology 
for an observable as $\theta_E$. The differential of $\theta_E$ is 
\be
\frac{\Delta \theta_E}{\theta_E} = \frac{\Delta \varphi}{\varphi} 
+ \Delta \ln \frac{d_{SL}}{d_{OS}} \,, 
\ee
which reveals how cosmological information is encoded just in the second 
term to the right.
By considering the cosmology of the DGP model for instance \cite{LUE}, one 
finds that the second term is $\sim -0.06 (\Delta \Omega_{\rm 
M}/\Omega_{\rm M}) + \Delta H_0/H_0$ for a source and a lens located at $z 
= 1$ and $z 
=0.3$, respectively ($\Omega_{\rm M} \sim 0.3$ and $H_0$ are the present 
matter density and Hubble parameter, respectively). On 
considering the uncertainties on the cosmological parameters of the 
order of percent,  
this simple quantitative example shows how the corrections to the 
Einstein deflection angle we have found in Table I should be taken into 
account in the study of strong lensing by clusters.

We believe that results similar to what we have found here for gDE models, 
might occur for dDE scenarios as well, in which the non-asymptotically flat 
term is due to the non-perturbative clumping of DE into objects detached from 
the cosmological expansion. However, dDE models 
may be less predictive than gDE models: gDE contain the 
same number of parameters of $\Lambda$CDM, while dDE 
may need more.  Let us end on noting that some of the gDE models 
considered here may have serious theoretical issues \cite{BD,koyama2}, 
whose 
resolution clearly go beyond the present project. However, the main result 
in Eq. (\ref{dephi}) of this paper remains valid: models alternative to 
general relativity with a cosmological constant predict a correction to 
the Einstein 
angle, which can be used to distinguish $\Lambda$ from other DE models.

\vspace{.5cm}

{\bf Acknowledgements} We wish to thank Lauro Moscardini for many 
discussions and suggestions on galaxies, groups and clusters. 
We are grateful to Robert R. Caldwell and one of 
the anonymous referees for valuable comments on this project.
FF and MG are partially supported by INFN IS PD51; FF and and AG are 
partially supported by INFN IS BO11. The authors thank the Galileo Galilei 
Institute for Theoretical Physics for the
hospitality and the INFN for partial support during the developments of 
this project.


\begin{thebibliography}{99}
\bibitem{same}
For semplicity we consider here the object and the observer at the same 
distance from the lens.
\bibitem{nota}
Note that the light bending in SdS written in terms of $J/E$ (rather than 
$r_0$) contains $\Lambda$ because of the relation between $r_0$ and $J/E$ 
in \cite{impact}.
\bibitem{note_mass}
Note that this is not true for a massive particle. $\Lambda$ induces a 
force on a test mass, for example producing an additional shift in the 
perihelion precession \cite{wright,matteo}.
\bibitem{note}
In the models studied here this fourth term is $\propto \gamma_2 ( 1 - 
\alpha)$: for $\gamma_2 > 0$ it is an attractive (repulsive) force for 
$\alpha=1/2$ ($3/2$). It is zero for conformal gravity, $\alpha=1$.
\end{thebibliography}
\end{document}